\documentclass[12pt]{article}
\usepackage{epsf}
\usepackage{rotating}

\newcommand{\lsim}{\raisebox{-0.13cm}{~\shortstack{$<$ \\[-0.07cm] $\sim$}}~}
\newcommand{\gsim}{\raisebox{-0.13cm}{~\shortstack{$>$ \\[-0.07cm] $\sim$}}~}
\newcommand{\gau}{\tilde{\chi}}
\newcommand{\beq}{\begin{equation}}
\newcommand{\eeq}{\end{equation}}
\newcommand{\tgb}{{\rm tg}\beta}

\catcode`@=11
\def\citer{\@ifnextchar
[{\@tempswatrue\@citexr}{\@tempswafalse\@citexr[]}}

\def\@citexr[#1]#2{\if@filesw\immediate\write\@auxout{\string\citation{#2}}\fi
  \def\@citea{}\@cite{\@for\@citeb:=#2\do
    {\@citea\def\@citea{--\penalty\@m}\@ifundefined
       {b@\@citeb}{{\bf ?}\@warning
       {Citation `\@citeb' on page \thepage \space undefined}}%
\hbox{\csname b@\@citeb\endcsname}}}{#1}}
\catcode`@=12

\begin{document}

\vspace*{-2mm}

\def\thefootnote{\fnsymbol{footnote}}
\rightline{
        \begin{minipage}{4cm}
	DESY 00--192\\
	Edinburgh~2000/31\\
	PM/00--45\\
	PSI--PR--00--19\\
        hep-ph/0101083\hfill \\
        \end{minipage}        
}

\vspace*{5mm}

\begin{center}

{\large\sc Production of MSSM Higgs Bosons \\[0.3cm]
in Photon-Photon Collisions\footnote{Supported in part by the European
Union under contract HPRN-CT-2000-00149.}}

\vspace{1cm}

{\sc M.M.~M\"uhlleitner$^{1,2}$, M.~Kr\"amer$^3$, M.~Spira$^4$ \\
and P.M.~Zerwas$^1$}

\vspace{0.5cm}

$^1$ DESY, Deutsches Elektronen-Synchrotron, \\
D-22603 Hamburg, Germany

\vspace*{0.3cm}

$^2$ Laboratoire de Physique Math\'ematique et Th\'eorique, UMR5825--CNRS, \\
Universit\'e de Montpellier II, F--34095 Montpellier Cedex 5,
France\footnote{present address}

\vspace*{0.3cm}

$^3$ Department~of Physics and Astronomy, University~of~Edinburgh, \\
Edinburgh EH9 3JZ, Scotland

\vspace*{0.3cm}

$^4$ Paul-Scherrer-Institut, \\
CH--5232 Villigen PSI, Switzerland

\end{center}

\vspace*{0.3cm}

\begin{abstract}
\noindent
{\it
  The heavy neutral Higgs bosons $H,A$ in the minimal supersymmetric extension
  of the Standard Model can be produced as single resonances at high-energy
  $\gamma\gamma$ colliders. We have studied the prospects of the search for
  these particles in $b\bar b$ and neutralino-pair final states. The
  Higgs bosons can be found with masses up to
  70--80\% of the initial $e^\pm e^-$ collider energy for medium
  values of $\tgb$, {\it i.e.} in areas of the supersymmetric parameter space
  not accessible at other colliders.
}
\end{abstract}

\def\thefootnote{\arabic{footnote}}

\section{Introduction}
The search for Higgs bosons \cite{higgs} is one of the most important
endeavours of present and future experiments \cite{1}. The minimal
supersymmetric extension of the Standard Model [MSSM] includes two
isodoublet Higgs fields which materialize, after electroweak symmetry
breaking, in five elementary Higgs bosons \cite{habergun}: two neutral
CP-even ($h,H$), one neutral CP-odd ($A$) and two charged ($H^\pm$)
particles.  To leading order, the MSSM Higgs sector can be described
by two parameters which are in general chosen as the pseudoscalar mass
$M_A$ and $\tgb = v_2/v_1$, the ratio of the two vacuum expectation
values of the scalar Higgs fields.  While the mass of the light
Higgs boson $h$ is bounded to $M_h\lsim 130$~GeV \cite{mssmrad}, the
masses of the heavy Higgs bosons $H,A,H^\pm$ are expected to be of the
order of the electroweak scale up to about 1 TeV. The heavy Higgs
bosons are nearly mass degenerate, $M_H-M_A \approx
(M_Z^2\sin^22\beta+\epsilon\cos^2\beta)/2M_A
\lsim 2$~GeV for $M_A \gsim 250$ GeV
with $\epsilon=3G_Fm_t^4 \log(1+M_{susy}^2/m_t^2)/\sqrt{2}
\pi^2 \sin^2\beta$. An important property of
the SUSY Higgs bosons is the
enhancement of the bottom Yukawa couplings with increasing $\tgb$.
Moreover, the couplings to supersymmetric particles, charginos and
neutralinos can be significant \cite{4}.

The light Higgs boson $h$ of the MSSM can be discovered at existing
or future $p\bar p/pp$ and $e^+e^-$ colliders. However, the heavy
Higgs bosons $H,A$ may escape detection in a wedge centered around
medium values of $\tgb\sim 7$,
and masses above 200 GeV even at the LHC
\cite{lhc}. At $e^\pm e^-$ linear colliders, heavy MSSM Higgs bosons can
only be discovered in associated production $e^+e^- \to HA$ \cite{3b}
according to the decoupling theorem. In first-phase $e^+e^-$
colliders with a total energy of 500 GeV, the heavy Higgs bosons can
thus be discovered with masses up to about 250 GeV. To extend the mass
reach, the $\gamma\gamma$ option of linear colliders can be used
in which high-energy photon beams are generated by Compton
back-scattering of laser light \cite{plc}. Neutral Higgs particles can be
formed as resonances in $\gamma\gamma$ collisions \cite{borden}:
\beq
\gamma\gamma \to h,H~\mbox{and}~A
\eeq
Center-of-mass energies of about 80\% of the primary $e^\pm e^-$ collider
energies and high
degrees of longitudinal photon polarization can be reached at
$\gamma\gamma$ colliders. Sufficient event rates are guaranteed for the
expected high-energy luminosities $\int {\cal L} =
300$~fb$^{-1}$ {\it per annum} \cite{9a}.  Photon
colliders therefore provide a useful instrument for the search for
heavy Higgs bosons not accessible elsewhere \cite{muehldiss}.

\section{Analysis}
\underline{\it Branching Ratios.}
As promising examples for the discovery of heavy MSSM Higgs bosons in
$\gamma\gamma$ collisions the two decay modes $H,A \to b\bar b$ and
$\gau^0 \gau^0$ have been investigated in detail for $\tgb=7$ and Higgs masses
$M_{H,A} > 200$ GeV. In this region the LHC is blind, if SUSY parameters
are not realized in a small favourable region, and linear colliders cannot
reach masses beyond 250 GeV in the first phase of the $e^+e^-$ mode. For
illustration, the additional MSSM parameters of the gaugino sector have
been chosen as $M_2=\pm\mu=200$~GeV, assuming a universal gaugino mass
at the GUT scale. These parameters correspond for positive $\mu$ to
neutralino masses $m_{\gau^0_{1,2,3,4}}=85, 148,
208, 271$~GeV and to chargino masses $m_{\gau^\pm_{1,2}} = 141, 270$~GeV.
For the sake of simplicity, sleptons and squarks are assumed to be so
heavy that they do not affect the analysis in a significant way.

\begin{figure}[hbtp]
\vspace*{-1.0cm}
\begin{minipage}[t]{7.2cm} {
\hspace*{0.5cm}
\epsfxsize=6.5cm \epsfbox{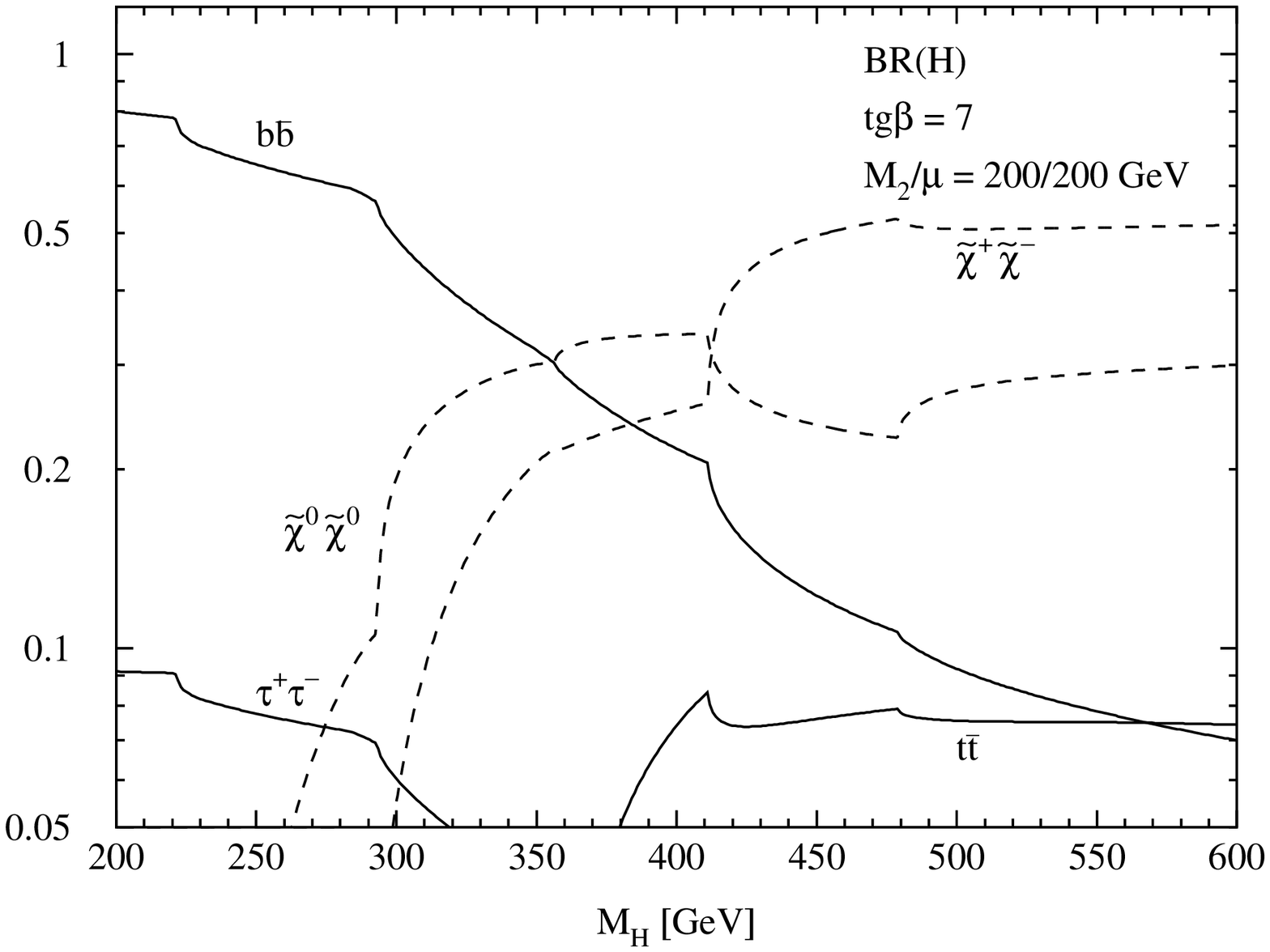}
}
\end{minipage}
\hspace*{1.0cm}
\begin{minipage}[t]{7.2cm} {
\hspace*{0.3cm}
\epsfxsize=6.5cm \epsfbox{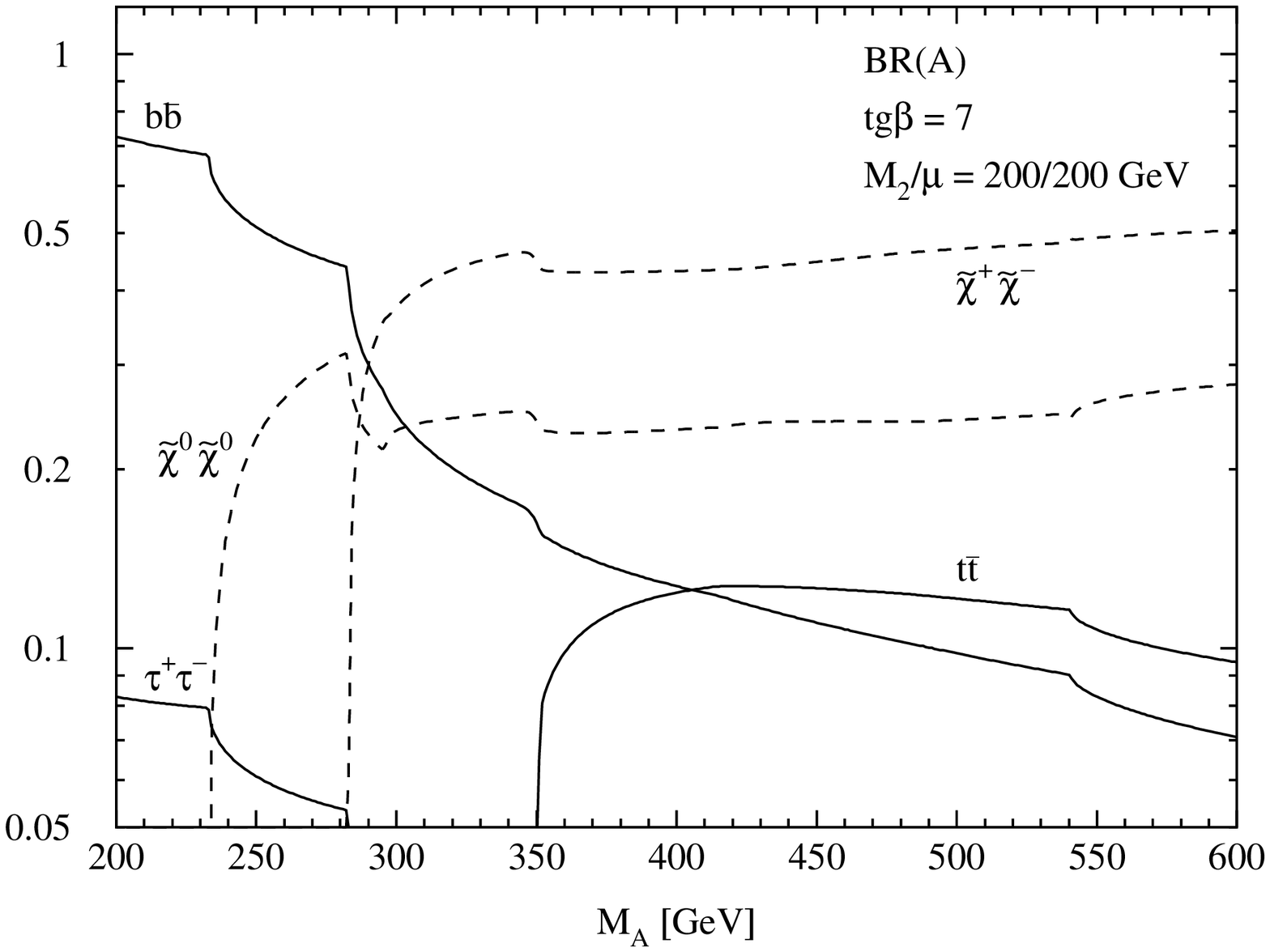}
}
\end{minipage}
\vspace*{-2.1cm}

\centerline{\bf (a)}

\vspace*{-1cm}
\begin{minipage}[t]{7.2cm} {
\hspace*{0.5cm}
\epsfxsize=6.5cm \epsfbox{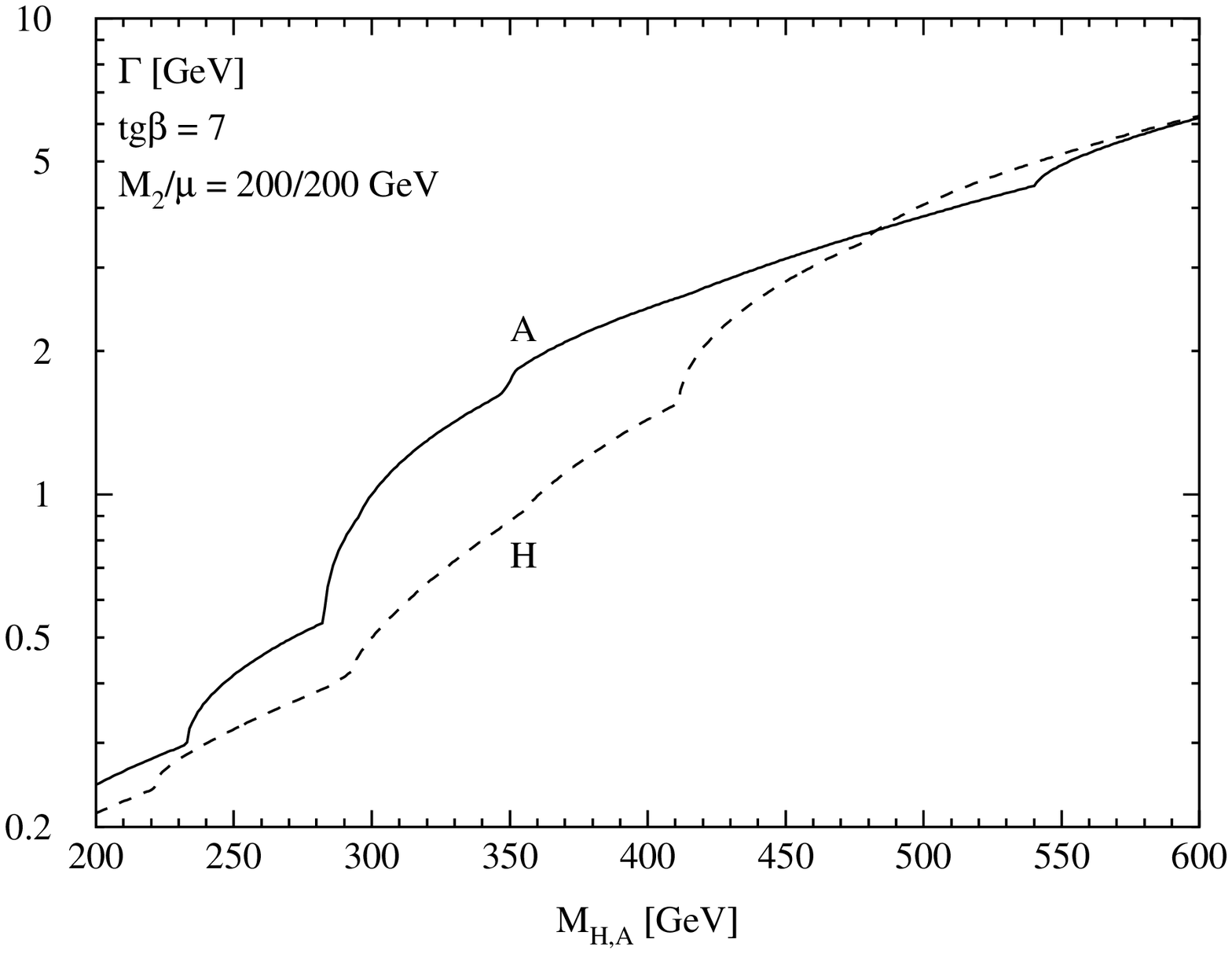}
}
\end{minipage}
\hspace*{1.0cm}
\begin{minipage}[t]{7.2cm} {
\hspace*{0.3cm}
\epsfxsize=6.5cm \epsfbox{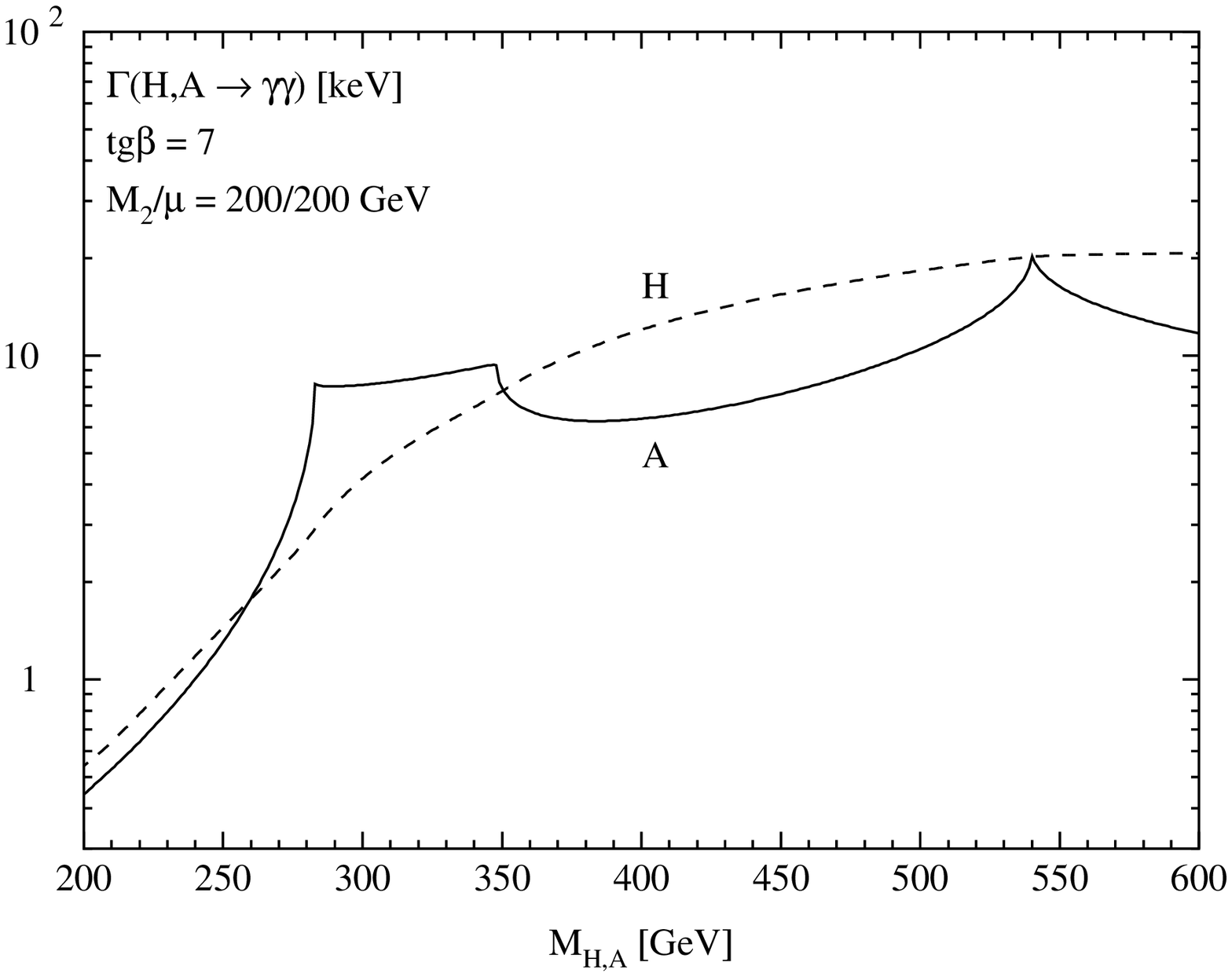}
}
\end{minipage}
\vspace*{-2.1cm}

\centerline{\bf (b)}

\caption[]{\it \label{fg:br} (a) Branching ratios, and (b) total widths
and partial $\gamma\gamma$ widths  of the heavy Higgs bosons
$H,A$ as a function of the Higgs masses. The MSSM parameters
have been chosen as $\tgb=7, M_2=\mu=200$ GeV.
The branching ratios are given for the sum of all charged 
and neutral $\gau\gau$ pairs except the LSP pair.}
\end{figure}

The decay branching ratios \cite{4,4c,hbbqcd0,hdecay}
are presented in Fig.~\ref{fg:br}a, the kinks
corresponding to thresholds of new channels. For
moderate masses, the $b\bar b$ decay modes turn out to be
dominant, while for large Higgs masses the dominant modes are
the decays into charginos and neutralinos. The decays to $\tau^+\tau^-$
and $t\bar t$ pairs are suppressed with respect to $b\bar{b}$ decays by
about an order of magnitude.

The partial decay widths $H,A\to \gamma\gamma$ determine the production
cross sections. The $\gamma\gamma$ couplings of the Higgs bosons are
mediated by charged-particle loops \cite{loop}. The loops of
top quarks and charginos
are dominant for the MSSM parameters chosen above. Congruent with the
uncertainty principle, the charginos decouple asymptotically
with increasing chargino masses.
The partial $\gamma\gamma$ widths are shown in Fig.~\ref{fg:br}b.

\vspace*{5mm}
\underline{\it Signals and Backgrounds.}
To leading order and in the narrow-width approximation, the production of
Higgs bosons in $\gamma\gamma$ collisions is described by the cross section
\beq
\langle \sigma(\gamma\gamma\to H,A) \rangle = \frac{8\pi^2}{s} 
 \frac{\Gamma(H,A\to
 \gamma\gamma)}{M_{H,A}} 
 \frac{d{\cal L}^{\gamma\gamma}}{d\tau_{H,A}}
\eeq
where $d{\cal L}^{\gamma\gamma}/d\tau_{H,A}$ denotes the differential
$\gamma\gamma$ luminosity, normalized to unity, for $\tau_{H,A}=M_{H,A}^2/s$.
The next-to-leading order QCD
corrections consist of gluon insertions in the quark triangle
loops of the photonic Higgs couplings and gluon radiation in the Higgs 
decays to quark-antiquark ($b\bar b$) pairs. The
two-loop contributions to the Higgs-$\gamma\gamma$ couplings are of moderate
size, if the quark mass inside the triangle loop is chosen as the
running quark mass at a typical scale determined by the c.m.~energy of the
process \cite{hggqcd}. The QCD corrections to the Higgs
decays into bottom quarks generate large logarithms, which however can
be absorbed in the running Yukawa coupling at the scale of the
c.m.~energy \cite{hbbqcd0,hbbqcd}. After inserting the running quark masses,
the total QCD corrections increase the signal cross section by about
20--40\%.

The main background processes $\gamma\gamma\to b\bar b$ and
$\gau^+\gau^-$ are pure QED reactions in lowest order. Since the
signals are generated for equal photon helicities, this configuration
can be enhanced by choosing the same helicities for the incoming laser
photons and helicities for the electrons/positrons that are opposite to the
laser photons in the initial state \cite{plc,kuehn}; at the same
time the background processes are suppressed. The NLO corrections to
$\gamma\gamma \to b\bar{b}$ for polarized photon beams have been
calculated in Ref.~\cite{bkgqcd}. They are moderate for photons
of opposite helicities but large for
photons of equal helicity. The differential cross section for
photons of equal helicity is suppressed by a factor $m_b^2/s$ at lowest
order due to the helicity flip in the bottom-quark line. This suppression
is removed by gluon radiation and the size of the cross
section increases to order $\alpha_s$. Large
Sudakov and non-Sudakov logarithms due to soft gluon 
radiation and soft gluon and bottom-quark
exchange in the virtual corrections must be resummed \cite{resum}.
In order to suppress the gluon radiation we have
selected slim two-jet configurations in the final state as defined
within the Sterman--Weinberg criterion. If the radiated gluon energy
is larger than $\epsilon_g = 10\%$ of the total $\gamma\gamma$ 
invariant energy
and if, at the same time, the opening angle between all three partons in
the final state is larger
than $\theta_c=20^o$, the event is classified as three-jet event and rejected.
The contamination of $b\bar{b}$ final states by the process
$\gamma\gamma\to c\bar{c}$ can be kept under experimental control by
$b$ tagging.

Moreover, the interference between the signal and background mechanisms has
been taken into account properly. This part affects only configurations
with equal photon helicities in the initial state. We have calculated the
next-to-leading order QCD corrections of the interference terms to quark
final states including the resummation of the large (non-) Sudakov logarithms
[for details see Ref.~\cite{muehldiss}]. The QCD corrections to the
interference terms are large.

\section{Results}
\underline{\it $b\bar b$ Channel.}
We assume that a rough scan in the $\gamma\gamma$ energy will first be
performed, from which preliminary evidence for the observation
of a Higgs boson can be derived. Since the
luminosity spectrum of equal-helicity photons is strongly peaked
at about 80\% of the $e^\pm e^-$ c.m.~energy \cite{plc,kuehn}, the
maximum can be tuned in the second step to the value
of the pseudoscalar Higgs mass $M_A$ at which the signal peak in the
energy scan appeared. The scalar resonance $H$ at a small distance
$M_H-M_A\sim 1$ GeV nearby is included by the non-zero experimental
resolution at the same time. The analysis can be optimized by
final-state cuts. The background is strongly reduced by a cut in the
production angle of the bottom quarks, $|\cos\theta | < 0.5$,
while the signal is affected only moderately.
By collecting $b\bar b$ final states with a resolution in
the invariant mass $M_A \pm 3$ GeV, that can be expected 
at photon colliders \cite{schreiber}, the sensitivity to the
combined $A$ and $H$ resonance peaks above the background is strongly
increased.
\begin{figure}[hbtp]
\vspace*{-2.5cm}
\hspace*{3.2cm}
\epsfxsize=9cm \epsfbox{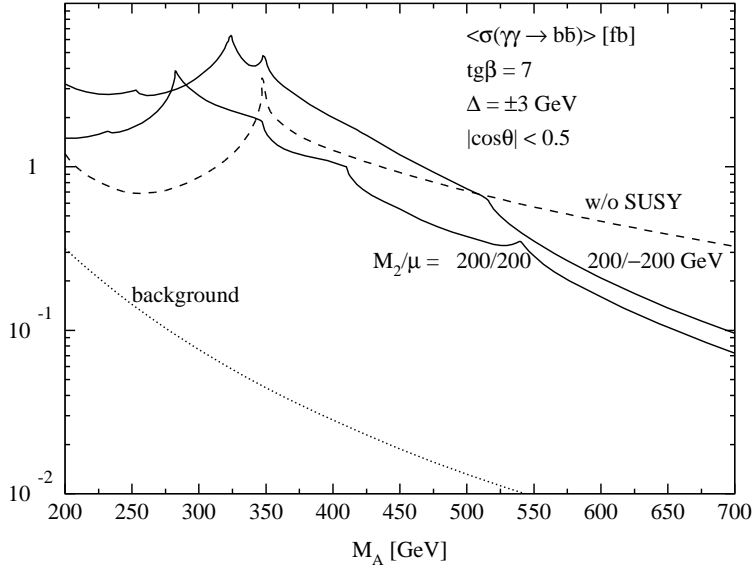}
\vspace*{-2.5cm}

\centerline{\bf (a)}
\vspace*{-1.5cm}

\hspace*{3.2cm}
\epsfxsize=9cm \epsfbox{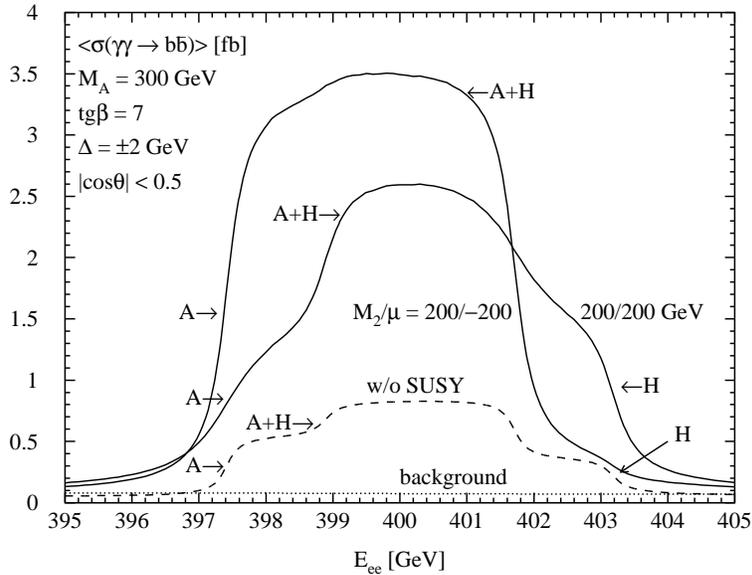}
\vspace*{-2.5cm}

\centerline{\bf (b)}
\vspace*{-0.0cm}

\caption[]{\it \label{fg:bot} (a) Cross sections for resonant heavy Higgs boson
  $H,A$ production in $\gamma\gamma$ collisions as a function of the
pseudoscalar Higgs mass $M_A$
  with final decays into $b\bar b$ pairs, and the corresponding
  background cross section. The maximum of the photon luminosity
  has been tuned to $M_A$.
  Cuts as indicated. The
  MSSM parameters have been chosen as $\tgb=7, M_2=\pm \mu=200$
  GeV; the limit of vanishing SUSY-particle contributions is shown 
  for comparison.
  (b) Threshold scans for
  $H,A$ production as a function of the $e^\pm e^-$
  collider energy with final decay into $b\bar b$ pairs. 
  The cross sections are defined
  in $b\bar b$ mass bins of $\pm~2~GeV$ around the maximum of the
  $\gamma\gamma$ luminosity.}
\end{figure}
The result for the peak cross section is shown in Fig.~\ref{fg:bot}a as
a function of the pseudoscalar mass $M_A$ for $\tgb=7$. It can be
inferred from the figure that the background is strongly suppressed.
The significance of the heavy Higgs boson signals is sufficient for the
discovery of the Higgs particles up to about 70--80\% of the $e^+e^-$
c.m.~energy. Thus, at a 500 GeV $e^+e^-$ linear collider the
$H,A$ bosons with masses up to about 400 GeV can be discovered in the
$b\bar b$ channel in the photon-collider mode, while for c.m.~energies
above 800 GeV the range can be extended to about 600
GeV. For heavier Higgs masses the signal rate
becomes too small for detection.

In a fine scan of the resonance region within a minimal bracket of
$\pm2$ GeV of the invariant-mass
resolution, the two Higgs bosons $A$ and $H$ can be
disentangled at least in part of the supersymmetry parameter space.
Increasing the energy stepwise from below, the Higgs
boson $A$ is produced first, followed by the combination of $A$ and $H$,
while finally $H$ is left before the scan leaves the resonance region.
This procedure is analyzed quantitatively in Fig.~\ref{fg:bot}b.
If the supersymmetry parameters are favourable, the
steps in the resonance formation curve are clearly visible.
However, this theoretical analysis must be backed by future experimental
simulations.

\vspace*{5mm}
\underline{\it $\gau^0\gau^0$ Channels.}
For the MSSM parameters introduced above, the heavy Higgs bosons have
significant decay branching ratios to pairs of charginos and
neutralinos, cf.~Figs.~\ref{fg:br}. However, due to the integer
chargino charge, the chargino background from continuum production is
in general more than an order of magnitude larger than the signal.
Pairs of neutralinos cannot be produced in $\gamma\gamma$ collisions
at leading order so that neutralino decays open a potential discovery
channel for the heavy Higgs bosons $H,A$ [see also Ref.~\cite{belan}].
This is apparent for moderate Higgs masses below the chargino decay
threshold. Detailed analyses of the topologies in the final state are
needed however to separate the neutralinos from the background
charginos above the chargino threshold. For example, the signal channel
$\gamma\gamma\to H,A\to \gau_2^0 \gau_1^0$ leads in the $\gau^0_2$
cascade decay predominantly to the hadronic final states
$jj + \not \!\! E$ with the jet-pair invariant mass
$M_{jj}$ clustering near the $\gau_2^0 - \gau_1^0$ mass
difference if the sfermions are heavy. By contrast,
the continuum process $\gamma\gamma\to\gau^+_1\gau^-_1$ generates the
final state $W^*W^*+\not \!\!\! E$ with distinct
$jjjj + \not \!\!\! E$ jet topologies. Thus, the neutralino decays
are expected to provide novel discovery channels of the heavy Higgs
bosons at photon-photon colliders.
\begin{figure}[hbtp]
\vspace*{-1.8cm}
\hspace*{3.0cm}
\epsfxsize=9cm \epsfbox{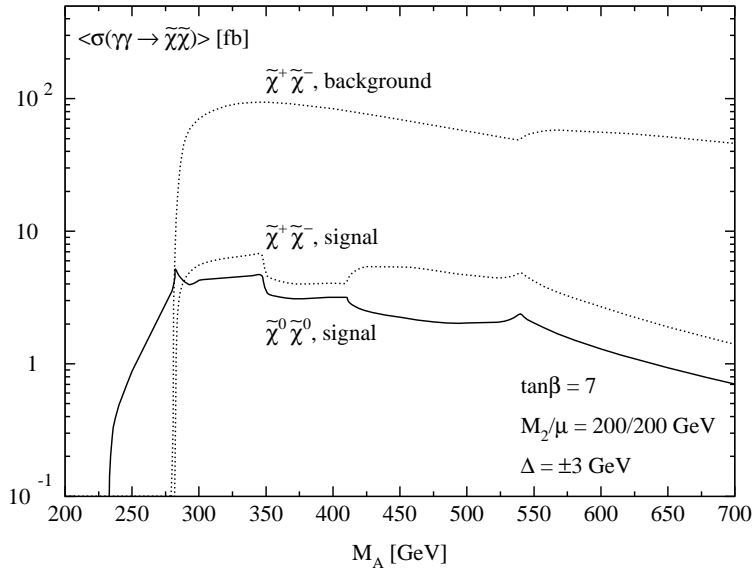}
\vspace*{-3.0cm}
\caption[]{\it \label{fg:gau} Same as Fig.~\ref{fg:bot}a for
  chargino and neutralino final states; $M_2=\mu=200$~GeV.}
\end{figure}

\section{Summary}
It has been shown in this letter that the heavy scalar and pseudoscalar
Higgs bosons $H$ and $A$ of the minimal supersymmetric extension of
the Standard Model MSSM can be discovered for medium values of $\tgb$
up to masses of about 400~GeV in the photon-photon collider mode
of a linear collider project in the first phase;
the mass reach can be extended to values
above 600 GeV at a TeV collider. The discovery potential for the heavy
Higgs bosons is unique since this region is not
accessible neither
in the respective $e^+e^-$ phase of a linear collider, nor at the LHC in
general.\\

\noindent
{\bf Acknowledgements.}
We are grateful to G.~Blair, A.~Djouadi, G.~Jikia, M.~Melles and
S.~S\"oldner-Rembold for
discussions and advice. MMM would like to thank the Particle Physics Theory 
Group at the University of Edinburgh for hospitality and PPARC for financial 
support.

\end{document}